\documentclass[12pt,journal=jctcce,manuscript=article,layout=onecolumn]{achemso}
\usepackage[latin9]{inputenc}
\usepackage{refstyle}
\usepackage{amsmath}
\usepackage{graphicx}
\PassOptionsToPackage{version=3}{mhchem}
\usepackage{mhchem}
\usepackage[numbers]{natbib}

\makeatletter

\title{Accurate core-excited states via inclusion of core triple excitations
in similarity-transformed EOM theory}

\author{Megan Simons}

\email{msimons@smu.edu}

\author{Devin A. Matthews}

\affiliation{Department of Chemistry, Southern Methodist University, Dallas, TX
75275}

\AtBeginDocument{\providecommand\figref[1]{\ref{fig:#1}}}
\RS@ifundefined{subsecref}
  {\newref{subsec}{name = \RSsectxt}}
  {}
\RS@ifundefined{thmref}
  {\def\RSthmtxt{theorem~}\newref{thm}{name = \RSthmtxt}}
  {}
\RS@ifundefined{lemref}
  {\def\RSlemtxt{lemma~}\newref{lem}{name = \RSlemtxt}}
  {}

\usepackage{siunitx}
\usepackage{mhchem}
\usepackage{pgfplots}
\usepackage[english]{babel}
\usepackage{amsmath}
\usepackage{amssymb}
\usepackage{stmaryrd}
\usepackage{xfrac}

\renewcommand{\figref}{\Figref}

\pgfplotsset{compat=1.3}
\usetikzlibrary{patterns}
\usetikzlibrary{plotmarks}

\@ifundefined{showcaptionsetup}{}{%
 \PassOptionsToPackage{caption=false}{subfig}}
\usepackage{subfig}
\makeatother

\begin{document}
\begin{abstract}
The phenomenon of orbital relaxation upon excitation of core electrons
is a major problem in the linear-response treatment of core-hole spectroscopies.
Rather than addressing relaxation through direct dynamical correlation
of the excited state via equation-of-motion coupled cluster theory
(EOMEE-CC), we extend the alternative similarity-transformed equation-of-motion
coupled cluster theory (STEOMEE-CC) by including the core-valence
separation (CVS) and correlation of triple excitations only within
the calculation of core ionization energies. This new method, CVS-STEOMEE-CCSD+cT,
significantly improves on CVS-EOMEE-CCSD and unmodified CVS-STEOMEE-CCSD
when compared to full CVS-EOM-CCSDT for K-edge core-excitation energies
of a set of small molecules. The improvement in both absolute and
relative (shifted) peak positions is nearly as good as for transition-potential
coupled cluster (TP-CC), which includes an explicit treatment of orbital
relaxation, and CVS-EOMEE-CCSD{*}, which includes a perturbative treatment
of triple excitations.
\end{abstract}

\section{Introduction}

Coupled cluster (CC) theory is one of the most powerful methods for
treating dynamical correlation in molecules, and is capable of computing
highly accurate ground state energies of small molecules.\citep{bartlettCoupledclusterTheoryQuantum2007}
Beyond the ground state, coupled cluster has been extended to excited
states via the time-dependent linear-response formalism,\citep{mukherjeeResponsefunctionApproachDirect1979,kochExcitationEnergiesCoupled1990}
the closely related equation-of-motion approach,\citep{geertsenEquationofmotionCoupledclusterMethod1989,comeauEquationofmotionCoupledclusterMethod1993,stantonEquationMotionCoupledcluster1993}
as well as other approaches such as the symmetry-adapted cluster technique.\citep{nakatsujiClusterExpansionWavefunction1978,nakatsujiClusterExpansionWavefunction1978a}
Equation-of-motion coupled-cluster (EOM-CC) theory can be used for
excited (EE-EOM-CC), electron-attached (EA-EOM-CC), and ionized state
(IP-EOM-CC) energies, as well as multiply-ionized/attached states
and even spin-flip excitations (SF-EOM-CC).\citep{levchenkoEquationofmotionSpinflipCoupledcluster2004,krylovSpinFlipEquationofMotionCoupledCluster2006}
EOM-CC is able to do this by taking advantage of the similarity transformation
of the Hamiltonian, which guarantees size-extensivity of the excited
state total energy (although excitation energies are not in general
size-consistent with respect to charge separation). While the ground
state CC wavefunction is single reference, the CI-like nature of the
EOM-CC wavefunction can capture significant multi-reference character
of excited states.\citep{krylovEquationofMotionCoupledClusterMethods2008,musialMultireferenceCoupledclusterTheory2011}

Nooijen et al. proposed an alternative approach to EOM-CC of using
a second similarity transformation of the Hamiltonian, followed by
diagonalization in a small (CIS-like) excitation space.\citep{nooijenNewMethodExcited1997,NooijenSimilaritytransformedequationofmotion1997}
In this similarity-transformed equation of motion coupled cluster
(STEOM-CC) theory, the second similarity transformation simultaneously
captures the dynamical correlation of all low-lying excited states.
In comparison, EOM-CC determines the wavefunction for a single excited
state and incorporates dynamical correlation effects via explicit
inclusion of higher excitations in the excited state wavefunction.
The STEOM-CC similarity transformation uses the wavefunctions of a
number of ionized and electron-attached states in order to build the
transformation. This transformation can be viewed as decoupling the
``active'' single-electron excitations from the double excitations,
much as the ground state coupled cluster equations decouple the reference
from single and double excitations. Thus, while STEOM-CC obtains the
excited state energies by diagonalization only in the space of single
excitations, it achieves an accuracy much greater than that of CIS
or EOM-CCS.

Both EOMEE-CCSD and STEOMEE-CCSD have been highly successful at describing
valence excited states of predominately single-excitation character,
but some modifications to the theory are necessary for application
in the x-ray regime. X-ray excited states, necessary for simulating
spectra such as NEXAFS, XES, and RIXS,\citep{normanSimulatingXraySpectroscopies2018}
are not bound states, but are resonances embedded deep in the valence
continuum. For EOM-CC, the core-valence separation (CVS) approach
of Coriani and Koch\citep{corianiCommunicationXrayAbsorption2015}
has been highly successful, although it is also possible to employ
other techniques such as damped response,\citep{corianiCoupledclusterResponseTheory2012}
complex scaling/complex absorption potential,\citep{zuevComplexAbsorbingPotentials2014}
etc. However, the CVS does not address the other major issue encountered
in the x-ray regime: orbital relaxation. Due to the presence of a
core hole in the excited state wavefunction, the valence orbitals
undergo considerable contraction and rotation. CVS-EOMEE-CCSD incompletely
captures this effect and hence overestimates core vertical excitation
energies by 1--3 eV (\emph{vide infra}). The addition of triple excitations,
either full CVS-EOMEE-CCSDT\citep{734c7d7e72234a548c61e39e075d7bd2}
or an approximate treatment of triples,\citep{matthewsEOMCCMethodsApproximate2020}
is necessary to fully treat the relaxation effects within standard
EOM-CC theory.

The CVS can be ported from EOM-CC to STEOM-CC in a relatively straightforward
manner.\citep{rangaCoreValenceSeparated2021} However, STEOM-CC also
offers an alternative approach to the orbital relaxation issue. Here
we propose a modification of CVS-STEOMEE-CCSD which efficiently and
accurate treats orbital relaxation for core excited states, which
we dub CVS-STEOMEE-CCSD+cT. This method compares favorably with standard
CVS-(ST)EOM-CCSD, and achieves similar performance compared to an
approximate inclusion of triples (CVS-EOMEE-CCSD{*}), as well as to
an explicit inclusion of relaxation effects in the reference via TP-CCSD.\citep{simonsTransitionpotentialCoupledCluster2021}

\section{Theoretical Methods}

\subsection{EOM-CC}

The coupled cluster ground state\citep{shavittManyBodyMethodsChemistry2009}
is characterized by a non-hermitian similarity transformation of the
Hamiltonian which decouples the reference from the space of excited
determinants,
\begin{align}
\bar{H} & =e^{-\hat{T}}He^{\hat{T}},\\
\hat{T}=\sum_{k=1}^{N}\hat{T}_{k} & =\sum_{ai}t_{i}^{a}a_{a}^{\dagger}a_{i}+\frac{1}{4}\sum_{abij}t_{ij}^{ab}a_{a}^{\dagger}a_{b}^{\dagger}a_{j}a_{i}+\cdots,\\
\langle\Phi_{0}|\bar{H}|\Phi_{0}\rangle & =E_{CC},\\
\langle\Phi_{i_{1}\ldots i_{k}}^{a_{1}\ldots a_{k}}|\bar{H}|\Phi_{0}\rangle & =0,\quad0<k\le N,
\end{align}
where $|\Phi_{0}\rangle$ is a zeroth order description of the wave
function (single determinant) and $|\Phi_{i_{1}\ldots i_{k}}^{a_{1}\ldots a_{k}}\rangle$
are the $k$th excited determinants. In EOM-CCSD, the CC equations
are solved in the space of single and double excitations and hence
$N=2$.

The transformed Hamiltonian $\bar{H}$, defined via the cluster operators
$\hat{T}_{k}$, provides the means to obtain excited states as well.
The ground state coupled cluster energy is an eigenvalue of $\bar{H}$
with distinct right and left eigenfunctions due to the non-hermitian
nature of the similarity transformation,
\begin{align}
\bar{H}\hat{R}(0)|\Phi_{0}\rangle & =E_{CC}\hat{R}(0)|\Phi_{0}\rangle,\quad\hat{R}(0)=\hat{1},\\
\langle\Phi_{0}|\hat{L}(0)\bar{H} & =\langle\Phi_{0}|\hat{L}(0)E_{CC},\quad\hat{L}(0)=\hat{1}+\hat{\Lambda}.
\end{align}

Explicit diagonalization of the transformed Hamiltonian, shifted by
the ground state energy, yields the vertical excitation energies $\omega_{i}$
and their associated right and left eigenfunctions,\citep{stantonEquationMotionCoupledcluster1993}
\begin{align}
\left(\bar{H}-E_{CC}\right)\hat{R}(m)|\Phi_{0}\rangle & =[\bar{H},\hat{R}(m)]|\Phi_{0}\rangle=\omega_{m}\hat{R}(m)|\Phi_{0}\rangle,\\
\hat{R}(m)=\sum_{k=0}^{N}\hat{R}_{k}(m) & =r_{0}(m)+\sum_{ai}r_{i}^{a}(m)a_{a}^{\dagger}a_{i}+\frac{1}{4}\sum_{abij}r_{ij}^{ab}(m)a_{a}^{\dagger}a_{b}^{\dagger}a_{j}a_{i}+\cdots,\\
\langle\Phi_{0}|\hat{L}(m)(\bar{H}-E_{CC}) & =\langle\Phi_{0}|\hat{L}(m)\omega_{m},\\
\hat{L}(m)=\sum_{k=1}^{N}\hat{L}_{k}(m) & =\sum_{ai}l_{a}^{i}(m)a_{i}^{\dagger}a_{a}+\frac{1}{4}\sum_{abij}l_{ab}^{ij}(m)a_{i}^{\dagger}a_{j}^{\dagger}a_{b}a_{a}+\cdots,
\end{align}
Note that the left-hand eigenfunction equations are not explicitly
connected. The connectedness of the right eigenfunction equations
allows for a solution purely in terms of $\hat{R}_{1}$ and $\hat{R}_{2}$.
Only one set of eigenfunctions is necessary to obtain the energy,
although both are necessary in order to calculate properties (including
transition properties) and analytic gradients.\citep{stantonAnalyticEnergyGradients1994}
We will use $\hat{R}(m)/\hat{L}(m)$ to refer to generic EOM-CC excitation/de-excitation
operators, or a subscript $EE$ to refer specifically to the EOMEE-CC
amplitudes. STEOMEE-CC requires, in addition, singly ionized and electron-attached
states which are formed via the application of non-number-conserving
excitation operators (here only for the right-hand side),
\begin{align}
\hat{R}_{IP}(m)=\sum_{k=1}^{N}\hat{R}_{IP;k}(m) & =\sum_{i}r_{i}(m)a_{i}+\frac{1}{2}\sum_{aij}r_{ij}^{a}(m)a_{a}^{\dagger}a_{j}a_{i}+\cdots\\
\hat{R}_{EA}(m)=\sum_{k=1}^{N}\hat{R}_{EA;k}(m) & =\sum_{a}r^{a}(m)a_{a}^{\dagger}+\frac{1}{2}\sum_{abi}r_{i}^{ab}(m)a_{a}^{\dagger}a_{b}^{\dagger}a_{i}+\cdots
\end{align}

\subsection{STEOM-CC}

Similarity transformed equation-of-motion coupled cluster theory\citep{NooijenSimilaritytransformedequationofmotion1997}
starts with the definition of the transformed Hamiltonian, $\hat{G}$,
\begin{align}
\{e^{\hat{S}}\}\hat{G} & =\bar{H}\{e^{\hat{S}}\},\\
\hat{G} & =\sum_{pq}g_{q}^{p}a_{p}^{\dagger}a_{q}+\frac{1}{4}\sum_{pqrs}g_{rs}^{pq}a_{p}^{\dagger}a_{q}^{\dagger}a_{s}a_{r}+\cdots
\end{align}
where the braces denote operator normal ordering (essentially, this
ensures that $\{e^{\hat{S}}\}$ has no internal contractions). The
transformation operator $\hat{S}$ has components in both the electron-attached,
or $(1,0)$, and ionized, or \textbf{$(0,1)$}, sectors of Fock space,
\begin{align}
\hat{S} & =\hat{S}^{+}+\hat{S}^{-},\\
\hat{S}^{+}=\sum_{k}\hat{S}_{k}^{+} & =\sum_{ae}s_{e}^{a}a_{a}^{\dagger}a_{e}+\frac{1}{2}\sum_{abej}s_{je}^{ba}a_{a}^{\dagger}a_{b}^{\dagger}a_{j}a_{e}+\cdots,\\
\hat{S}^{-}=\sum_{k}\hat{S}_{k}^{-} & =\sum_{im}s_{i}^{m}a_{m}^{\dagger}a_{i}+\frac{1}{2}\sum_{ijmb}s_{ji}^{bm}a_{m}^{\dagger}a_{b}^{\dagger}a_{j}a_{i}+\cdots,
\end{align}
where $e$ and $m$ are associated with sets of $n_{v;act}$ active
virtual and $n_{o;act}$ active occupied orbitals (transformed back
to the canonical MO space).

As noted by Nooijen,\citep{NooijenSimilaritytransformedequationofmotion1997}
the $\hat{S}_{1}$ amplitudes are not necessary for the solution of
the STEOMEE-CC equations as they simply cause rotations within the
singles excitation space, and hence a diagonalization within the \emph{full}
singles space is invariant. In fact, diagonalization within the full
rather than the active singles space is desirable as the portion of
the solution falling outside the active space serves as a measure
of active space insufficiency.\citep{NooijenSimilaritytransformedequationofmotion1997}

The transformation amplitudes $\hat{S}^{\pm}$ may be easily obtained
by renormalization of a set of solutions of the EOMIP-CC and EOMEA-CC
equations,
\begin{align}
S_{ji}^{bm} & =-\sum_{\kappa\lambda=1}^{n_{o;act}}r_{ji}^{b}(\lambda)\left(U_{-}^{-1}\right)_{\lambda\kappa}\delta_{\kappa m},\label{eq:s2-}\\
S_{je}^{ba} & =\sum_{\kappa\lambda=1}^{n_{v;act}}r_{j}^{ba}(\lambda)\left(U_{+}^{-1}\right)_{\lambda\kappa}\delta_{\kappa e},
\end{align}
where the extra minus sign for the IP-coefficients comes from the
contraction over the hole line $\lambda$ in (\ref{eq:s2-}). The
matrices $U_{\pm}$ are the transformation matrices which diagonalize
the STEOM effective Hamiltonian. They are derived from the single
excitation parts of the IP/EA solutions,
\begin{align}
\left(U_{-}\right)_{\kappa\lambda} & =\sum_{n}\delta_{\kappa n}r_{n}(\lambda),\\
\left(U_{+}\right)_{\kappa\lambda} & =\sum_{f}\delta_{\kappa f}r^{f}(\lambda).\label{eq:uplus}
\end{align}

In (\ref{eq:s2-})--(\ref{eq:uplus}), the factor $\delta_{\kappa p}$
indicates that the active orbitals (indexed by $\kappa$) are simply
a subset of the canonical molecular orbitals (indexed by $p$), typically
the orbitals within a small energy range around the Fermi level. The
solutions $\hat{R}_{IP/EA}(\lambda)$ typically correspond to principal
ionizations from and electron attachments to these active orbitals.
The active orbitals may also be chosen as linear combinations of molecular
orbitals,\citep{duttaAutomaticActiveSpace2017,rangaCoreValenceSeparated2021}
with the proper transformation matrix replacing the Kronecker delta.

The excited states are obtained by solving the eigenvalue equations,
\begin{equation}
[\hat{G},\hat{R}_{1}(m)]|\Phi_{0}\rangle=\omega_{m}\hat{R}_{1}(m)|\Phi_{0}\rangle.\label{eq:steom-eig}
\end{equation}
The left-hand eigenvalue equations formally require a solution in
the full singles and doubles space, although computation of properties
and transition strengths can be simplified by a perturbative approximation
of $\hat{L}_{2}$.\citep{NooijenSimilaritytransformedequationofmotion1997}

\subsection{Core Excited States}

The direct calculation of core excited states, even for the 1s orbitals
of first row elements with energies on the order of 100--500 eV,
is fraught with difficulties. Standard EOMEE-CC calculations are difficult
or impossible to converge, and even when convergence is achieved,
the energies may be contaminated by spurious couplings to high-lying
valence excitations.\citep{liuBenchmarkCalculationsKEdge2019}

As a remedy to this problem, Coriani and Koch\citep{corianiCommunicationXrayAbsorption2015}
adapted the core-valence separation scheme first introduced by Cederbaum,
Domcke, and Schirmer\citep{cederbaumManybodyTheoryCore1980} to EOM-CC.
In CVS-EOM-CC, amplitudes which correspond to purely valence excitations
are explicitly zeroed. This leaves only components involving one or
more core orbitals (indicated by capital roman letters),
\begin{equation}
\hat{R}_{CVS}=\sum_{aI}r_{I}^{a}(m)a_{a}^{\dagger}a_{I}+\frac{1}{2}\sum_{abIj}r_{Ij}^{ab}(m)a_{a}^{\dagger}a_{b}^{\dagger}a_{j}a_{I}+\frac{1}{4}\sum_{abIJ}r_{IJ}^{ab}(m)a_{a}^{\dagger}a_{b}^{\dagger}a_{J}a_{I}+\cdots.\label{eq:rcvs}
\end{equation}
This formulation eliminates all core-valence couplings and recovers
the core-excited states as bound solutions. A number of variations
on this basic theme exist, such as including only one or a small number
of symmetry-related core orbitals in the ``core'' set and treating
the rest as valence,\citep{734c7d7e72234a548c61e39e075d7bd2} solving
the ground state equations in the frozen-core approximation,\citep{vidalNewEfficientEquationofMotion2019}
etc. In this work we target a single core orbital in each calculation
and define $\hat{R}$ as in (\ref{eq:rcvs}), with an all-electron
solution of the ground state. Note that we purposefully avoid test
molecules with symmetric nuclei as these cases require either treating
the symmetric and anti-symmetric core molecular orbitals equally or
breaking the symmetry via orbital localization.

Within STEOMEE-CCSD, the CVS can be applied in two places, giving
rise to CVS-STEOMEE-CCSD. First, the selection of the active space
must include the core orbital(s) of interest, and when solving for
these core ionization wavefunctions the CVS is necessary to stabilize
convergence and provide solutions without spurious coupling. Second,
the solution of the eigenvalue equations (\ref{eq:steom-eig}) may
also use the CVS in defining the excitation operator $\hat{R}_{1}$
in order to accelerate the computation. Note that if only one core
orbital is included in the CVS treatment, then the diagonalization
step scales as only $\mathcal{O}(n_{root}n_{v}^{2})$, where $n_{v}$
is the number of virtual orbitals. The application of the CVS to the
$\hat{R}_{IP}$ operator does not induce a similar restriction on
the $\hat{S}^{-}$ operator. Because this operator contains information
from \emph{both} valence and core ionized wavefunctions, it must necessarily
span the full molecular orbital space. The combined valence and core
nature of $\hat{S}^{-}$ also allows for the simultaneous and balanced
determination of core and valence excitation energies.\footnote{The ground state, IP/EA, and $\hat{G}$ computations may be shared,
although separate diagonalization steps are still necessary as the
core and valence wavefunctions are not strictly orthogonal and have
different structure. The diagonalization(s) are computational very
inexpensive, however. } In principle it may be possible to construct an $\hat{S}^{-}$ and
hence $\hat{G}$ operator using \emph{only} core ionization wavefunctions
if no valence excitations are desired. We have not explored this possibility
as the valence IP solutions are rarely a bottleneck in practice.

The CVS eliminates most convergence issues and recovers excitation
energies which are systematically improvable towards the experimental
value with increasing basis set size and level of excitation.\citep{734c7d7e72234a548c61e39e075d7bd2,sarangiBasisSetSelection2020}
However, at the singles and doubles level, large errors remain which
are attributable to the significant orbital relaxation from the ground
to the excited core-hole state. We previously introduced the transition-potential
coupled cluster method (TP-CC),\citep{simonsTransitionpotentialCoupledCluster2021}
which accounts for orbital relaxation explicitly by performing the
calculation with molecular orbitals optimized for a fractional core-hole.
Instead, the STEOMEE-CC approach offers an alternative.

The ionization part of the similarity transformation, defined by the
solution of the standard EOMIP-CC equations, provides all of the necessary
dynamical correlation for the occupied orbitals. Then, since the inclusion
of triple excitations in EOMEE-CC provides a necessary level of correlation
to account for core-hole orbital relaxation, we propose that inclusion
of triple excitations, \emph{only in the EOMIP-CC solutions}, is sufficient
to account for orbital relaxation in STEOMEE-CC. This modification
introduces an additional transformation operator $\hat{S}_{3}^{-}$,
\[
S_{kji}^{cbm}=-\sum_{\kappa\lambda=1}^{n_{O;act}}r_{kji}^{cb}(\lambda)\left(U_{-}^{-1}\right)_{\lambda\kappa}\delta_{\kappa m},
\]
where at least one of $ijk$ must be an active core orbital, and $\kappa\lambda$
refer to the $n_{O;act}$ active core orbitals. Since STEOMEE-CC only
requires elements of $\hat{G}$ with at most two lines at the top
and bottom, diagrammatically, the only modifications necessary to
the formation of the twice-transformed Hamiltonian are,
\begin{align*}
g_{i}^{m} & \leftarrow S_{kji}^{cbm}\langle jk\Vert bc\rangle,\\
g_{ei}^{ma} & \leftarrow-S_{ijk}^{abm}\langle jk\Vert be\rangle.
\end{align*}
The diagonalization step proceeds exactly the same as in unmodified
CVS-STEOM-CCSD. We denote this new method as CVS-STEOMEE-CCSD+cT (``singles
and doubles plus core triples''). For a single core orbital, the
solution of the CVS-EOMIP-CCSDT equations scales as $\mathcal{O}(n_{o}^{2}n_{v}^{4})$,
which is the same as the ground state CCSD equations. Also note that
the connectivity of the EOMIP-CCSDT equations is preserved even without
including triple excitations in the ground state.

\section{Computational Details}

CVS-STEOMEE-CCSD and CVS-STEOMEE-CCSD+cT were implemented in a development
version of the CFOUR program package.\citep{matthewsCoupledclusterTechniquesComputational2020}
In all cases we included all canonical orbitals with orbital energies
between -20 and +10 eV as active orbitals. A single core orbital was
included in the CVS treatment and STEOM principal IP calculations
in each experiment. We observed only very small (a few 10s of meV)
changes upon expansion of the active space.

The test set consisted of four vertical core excitation energies from
each 1s core orbital of $\ce{H2O}$, CO, HCN, HF, HOF, HNO, $\ce{CH2}$,
$\ce{CH4}$, $\ce{NH3}$, $\ce{H3CF}$ $\ce{H3COH}$, $\ce{H2CO}$,
$\ce{H2CNH}$, and $\ce{H2NF}$. This includes a total of 94 vertical
excitation energies. The core excitations were selected as those for
which we could reliably converge all methods tested, which typically
consisted of the first four excitations of dominant single excitation
character. All calculations utilized the aug-cc-pCVTZ basis set with
all electrons correlated, except for $\ce{H2O}$ where aug-cc-pCVQZ
was used. In order to avoid complications due to missing relativistic
effects, basis set incompleteness (particularly for Rydberg core excitations),
geometric effects, and data quality and availability, which would
all be a concern when comparing directly to experimental data, we
have used full CVS-EOM-CCSDT as a benchmark, as in previous work.
\citep{simonsTransitionpotentialCoupledCluster2021}

\section{Results and Discussion}

\begin{figure}
\includegraphics[scale=0.75]{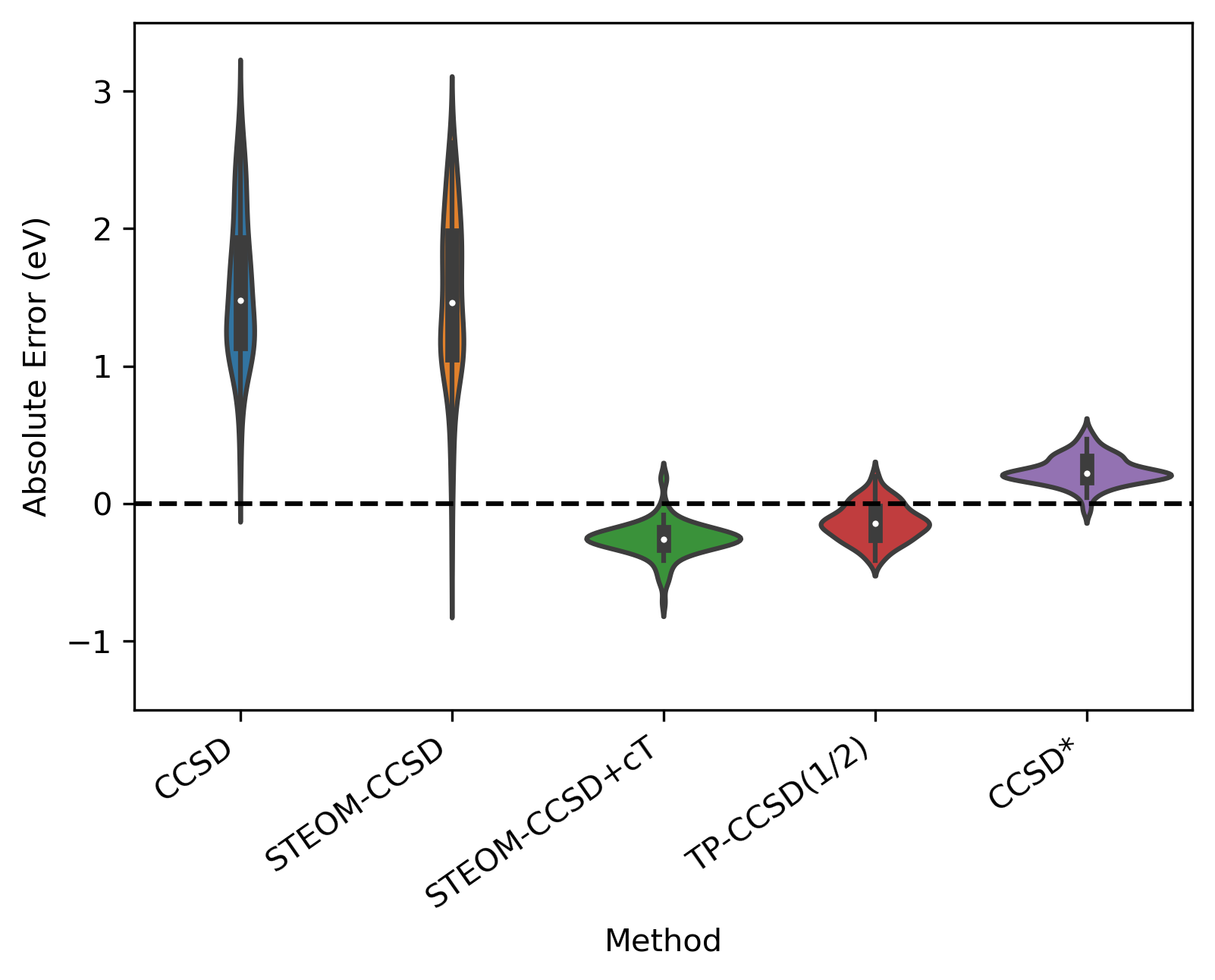}

\caption{\label{fig:absolute}Error distributions with respect to CVS-EOM-CCSDT
for absolute vertical core excitation energies. }
\end{figure}
\begin{figure}

\subfloat[\label{fig:rel-ex}]{

\includegraphics[scale=0.75]{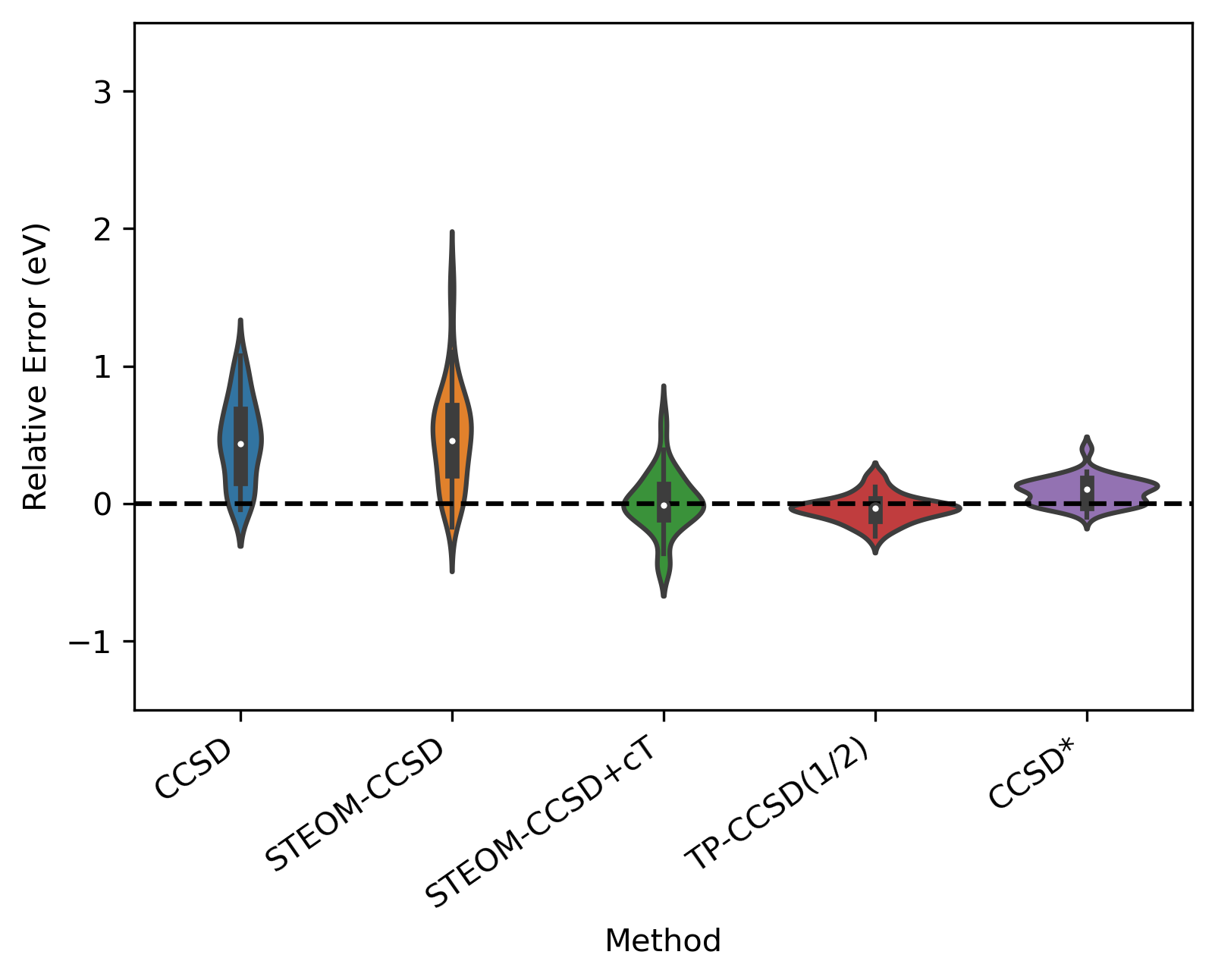}}

\subfloat[\label{fig:term}]{\includegraphics[scale=0.75]{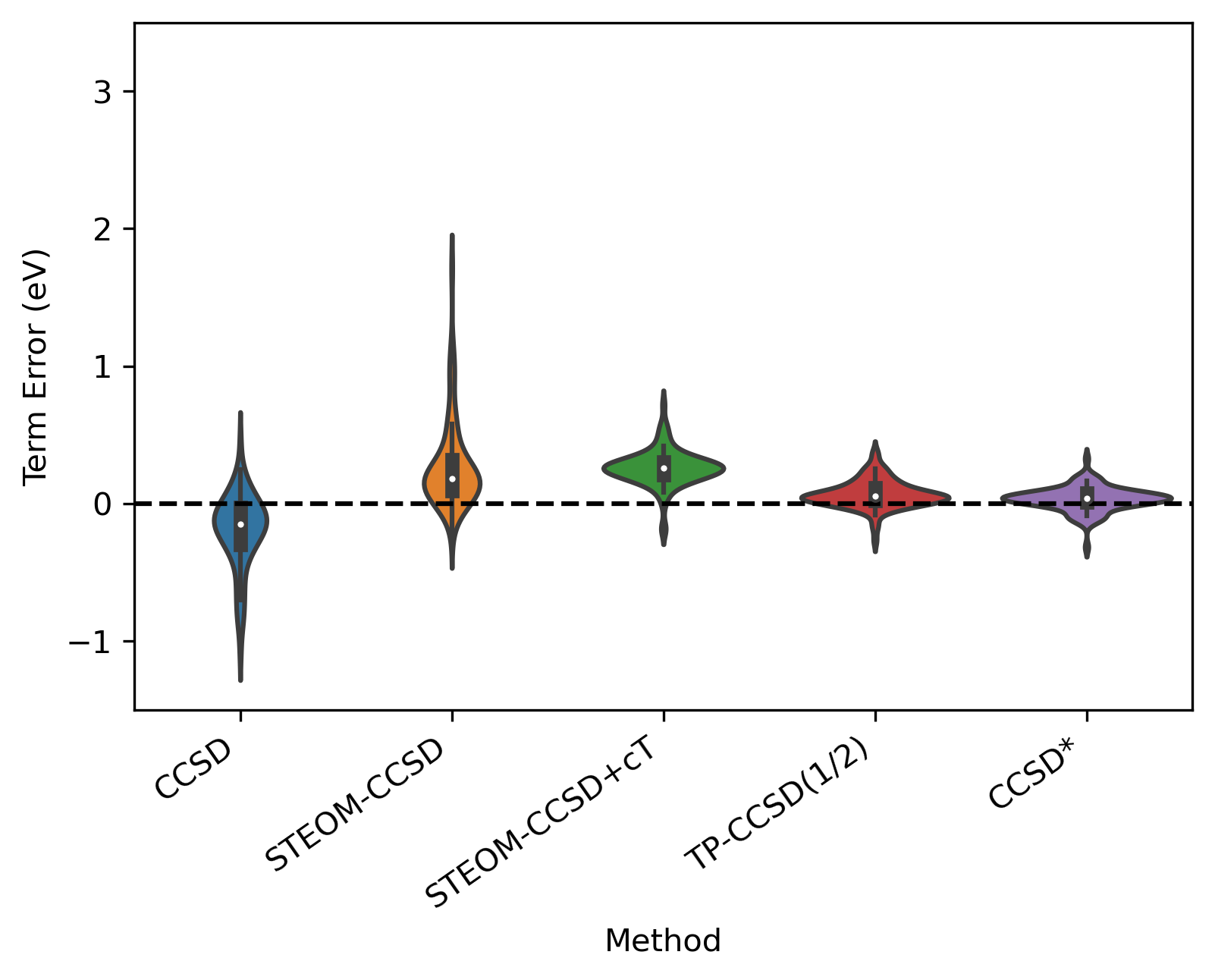}

}\caption{\label{fig:rel}Error distributions for relative vertical excitation
energy errors with respect to CVS-EOM-CCSDT. (a) Error in position
relative to lowest excitation energy within the pre-edge region, and
(b) error relative to the corresponding ionization edge.}
\end{figure}

In the following discussion and in Figs. \ref{fig:absolute} and \ref{fig:rel},
the ``shortened'' names of the CVS-EOM methods will be used, e.g.
CCSD = CVS-EOM-CCSD, with the exception of TP-CCSD(1/2). The distribution
of ``absolute'' (i.e. unmodified vertical) excitation energy deviations
from CCSDT are depicted in \figref{absolute}. The absolute energy
deviation for a method $X$ is calculated as $E(X)-E(\text{CCSDT})$
where $E$ is a vertical core excitation energy. The ``relative''
excitation energy deviations are depicted in \figref{rel}. These
deviations are determined in two ways: a) from excitation energies
measured relative to the lowest excitation energy in each edge $E_{rel}(X)=E(X)-E_{1}(X)$,
and b) from excitation energies measured downwards from the ionization
edge, $E_{rel}(X)=IP(X)-E(X)$, and then deviations are computed as
before. In the former case the first excitation energy (which has
zero relative error by definition) is not included in the distribution.
The use of the ionization edge or excitation onset as a reference
causes a shift of the entire spectrum which is method- and molecule-specific.
Note that the core ionization potentials for CVS-STEOMEE-CCSD are
identical to EOMIP-CCSD, while those for CVS-STEOMEE-CCSD+cT are the
same as ``EOMIP-CCSD(2,3)'', i.e. EOMIP-CC with a CCSD ground state
and an EOMIP-CCSDT ionized state.

From \figref{absolute}, it is clear that both ``purely singles and
doubles'' methods, CCSD and STEOM-CCSD are prone to large errors
in absolute vertical core excitation energies. In the case of CCSD,
errors are as large as 3 eV but uniformly positive, reflecting the
tendency of orbital relaxation to lower the excitation energies. STEOM-CCSD
errors cover a slightly larger range, with a few states predicted
too low, but overall the performance of the two methods is similar.
The remaining methods: TP-CCSD(1/2) (which includes explicit orbital
relaxation), CCSD{*} (which includes explicit triple excitations in
the excited state), and our new method STEOM-CCSD+cT, all improve
significantly, lowering absolute errors to less than 1 eV and less
than 0.5 eV in the typical case.

On the other hand, \figref{rel} represents a more useful picture
of error characteristics by including a shift of the spectrum for
each ionization edge. Since such as shift is almost always applied
when comparing to experimental spectra, these error distributions
reflect the remaining errors which directly affect the structure of
the pre-edge region. Large errors here (larger than the experimental
line widths of $\sim0.3$ eV) can lead to errors in assignment and
analysis of experimental spectra. Errors for CCSD and STEOM-CCSD are
indeed reduced, although observed deviations from CCSDT cover almost
1.5 eV for CCSD and 2.5 eV for STEOM-CCSD. Here we note that STEOM-CCSD
does indeed display noticeably worse performance compared to CCSD,
with the larger number of outliers (almost all valence $\pi^{*}$
states) significantly stretching the error distribution. This effect
is particularly striking in \figref{term} where the otherwise nicely
compact error distribution of STEOM-CCSD is ruined by large positive
errors for such valence states (here, a positive error indicates that
the valence states lie too low, or conversely that the Rydberg series
lies too high). TP-CCSD(1/2) and CCSD{*} essentially achieve the goal
of relative errors in the range of 0.3 eV for both types of relative
error.

Errors for STEOM-CCSD+cT relative to the ionization edge (\figref{term})
show a fairly compact distribution but an overall downward shift of
$\sim0.3$ eV for the entire excitation spectrum. This may be due
to imbalance in the level of correlation of the core ionization potential,
which now includes full triples, and the excitation spectrum, which
derives largely from the valence electron affinity calculation which
remain purely singles and doubles. An approximate treatment of triples
in the core ionization potential calculation would further speed up
the calculation and also possibly correct for this excessive gap.
When comparing to the lowest excitation energy (\figref{rel-ex}),
STEOM-CCSD+cT displays a compact distribution flanked by two outlying
wings. The positive wing is essentially entirely due to $\ce{CH2}$,
where the gap between the excitation into the lone pair and the Rydberg
states is overestimated by 0.6 eV. The negative wing is dominated
by the fluorine K-edges of HOF and $\ce{H2NF}$, which underestimate
the gap between the mixed $\sigma^{*}/3s$ excitation and the remaining
excitations by 0.4--0.5 eV. While these cases represent challenging
electronic structures for STEOM-CCSD+cT, the overall pre-edge structure
is maintained well except for these single gaps. Noticeably, the large
errors for $\pi^{*}$ valence states present in STEOM-CCSD are almost
entirely eliminated in STEOM-CCSD+cT due to the improved description
of the core hole.

\section{Conclusions and Future Work}

The problem of orbital relaxation is central to the accurate computation
of core-hole spectra. While our previous work focused on an explicit
inclusion of core relaxation via the use of orbitals optimized for
fractional core occupation, here we show that a simple modification
of similarity-transformed equation-of-motion theory can similarly
address this important problem. Using full CVS-EOM-CCSDT as a benchmark,
our new CVS-STEOMEE-CCSD+cT method reduces errors in absolute vertical
core excitation energies by a factor of $\sim5$ compared to CVS-EOMEE-CCSD,
and also improves errors in pre-shifted excitation spectra to less
than 0.5 eV, except for the challenging cases of $\ce{CH2}$ and fluorine
K-edges with valence-Rydberg mixing. The addition of triple excitations
only affects the core ionization calculation and scales as $\mathcal{O}(n_{c}n_{o}^{2}n_{v}^{4})$
for $n_{c}$ active core, $n_{o}$ occupied, and $n_{v}$ virtual
orbitals. This results in a modest increase in computational effort
compared to standard CVS-EOM-CCSD, since the solution of the electron-attachment
problem is typically the main bottleneck other than the ground state.

We are currently implementing CVS-STEOM-CC excited state and transition
properties, which require little additional development over the standard
theory.\citep{NooijenSimilaritytransformedequationofmotion1997,rangaCoreValenceSeparated2021}
Importantly, the inclusion of triple excitations in the core ionization
potential calculation does not directly enter the computation of these
properties. In another vein, the approximate inclusion of triple excitations
may offer an equally effective and even more computationally inexpensive
option. With the advances developed in this and future work, we are
confident that similarity-transformed equation-of-motion theory will
find significant use in core-hole spectroscopy, not only due to the
excellent accuracy exhibited in our benchmark, but also due to the
ability of STEOM to treat large numbers of excited states efficiently,
its ability to treat core and valence states on an even footing, and
its proven computational efficiency.

\section*{Acknowledgments}

The authors would like to thank Prof. Marcel Nooijen for inspiring
us to work on triple excitations in STEOM-CC theory and for several
fruitful discussions. This work was supported in part by the US National
Science Foundation under grant OAC-2003931. MS is supported by an
SMU Center for Research Computing Graduate Fellowship. All calculations
were performed on the ManeFrame II computing system at SMU.

\section*{Supplementary Material}

An electronic supplementary information file is available as an Excel
file (.xslx). This file contains the employed molecular geometries
and basis sets and the complete results, comprising absolute and relative
excitation energies, and ionization potentials.

\section*{Data Availability}

The data that supports the findings of this study are available within
the article and its supplementary material.

\bibliography{paper}
\pagebreak{}

\begin{figure}
\includegraphics[scale=0.75]{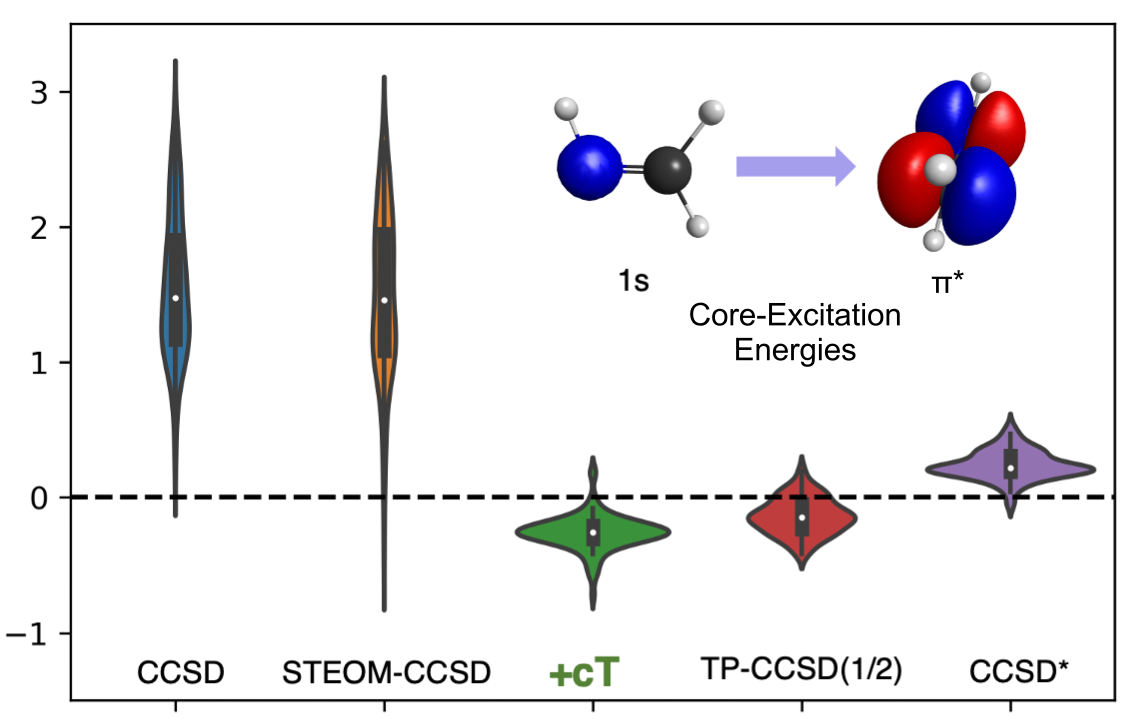}

\caption{For Table of Contents Only}

\end{figure}

\end{document}